%% file: paper.tex
\documentclass{sig-alternate}
\usepackage{verbatim} 
\usepackage{makeidx}  
\usepackage{amssymb}
\usepackage{bbm}
\usepackage{graphicx}
\usepackage{tikz}
\usetikzlibrary{arrows,backgrounds}
\usepackage{booktabs}

\newcommand\blfootnote[1]{%
  \begingroup
  \renewcommand\thefootnote{}\footnote{#1}%
  \addtocounter{footnote}{-1}%
  \endgroup
}

\usepackage{etoolbox}
\patchcmd{\maketitle}{\@copyrightspace}{}{}{}

\begin{document}
\title{
Intent Models for Contextualising and \\ Diversifying Query Suggestions
}
\numberofauthors{1}
\author{
\alignauthor
Eugene Kharitonov$^{\dagger \ddagger}$, Craig Macdonald$^\ddagger$, Pavel Serdyukov$^\dagger$, Iadh Ounis$^\ddagger$\\
       \affaddr{$^\dagger$Yandex $^\ddagger$ School of Computing Science, University of Glasgow} \\
    \email{$^\dagger$\{kharitonov, pavser\}@yandex-team.ru $^\ddagger$\{craig.macdonald, iadh.ounis\}@glasgow.ac.uk}
}

\maketitle
\blfootnote{A short version of this paper \cite{Kharitonov2013} was presented at ACM CIKM 2013.}

\begin{abstract}
\looseness-1 The query suggestion or auto-completion mechanisms help users to type less while interacting with a search engine. A basic approach that ranks suggestions according to their frequency in the query logs is suboptimal. Firstly, many candidate queries with the same prefix can be removed as redundant. Secondly, the suggestions can also be personalised based on the user's context. These two directions to improve the aforementioned mechanisms' quality can be in opposition: while the latter aims to promote suggestions that address search intents that a user is likely to have, the former aims to diversify the suggestions to cover as many intents as possible. We introduce a contextualisation framework that utilises a short-term context using the user's behaviour within the current search session, such as the previous query, the documents examined, and the candidate query suggestions that the user has discarded. This short-term context is used to contextualise and diversify the ranking of query suggestions, by modelling the user's information need as a mixture of intent-specific user models.
The evaluation is performed offline on a set of approximately 1.0M test user sessions. Our results suggest that the proposed approach significantly improves query suggestions compared to the baseline approach.
\end{abstract}

\section{Introduction}
Query suggestion\footnote{The term \emph{query auto-completion} \cite{Bar2011} is also used.} 
is a mechanism that helps a search engine's users to type less while submitting a query. 
Usually, such suggestions are represented as a list of queries, which are filtered by the prefix entered by the user. This list appears as a user starts to enter a new query and changes as the user types new characters.
 The most common approach to generate query suggestion candidates for web search is based on query log mining (e.g., \cite{Bar2011, Alisa2012}). 
Usually the process can be divided into offline preparation and online suggestion steps. In the offline step, queries are aggregated from search logs, cleaned from inappropriate/adult queries and clustered to identify near duplicates. Finally, the queries are indexed. In the online step, this index is used to provide a user with a small set of possible candidates (usually, no more than 10) immediately after the user starts to type the query. Since short prefixes can have thousands of highly frequent candidates, the problem is to select a subset that has a maximum probability to contain the query that the user is trying to submit.
Since up to 50\% of queries in the query stream are preceded by other queries within the same session, the user's earlier behaviour in the session can be a valuable source of contextual information \cite{Bar2011}, which can be used to provide the user with a better list of suggestions.  To illustrate the utility of such kind of context, 
let us consider a user has just entered the query [apache], clicked on several pages devoted to Native Americans and skipped \emph{http://www.apache.org/}. It is natural to assume that the user's next input [apache t] refers to [apache territory] and not to [apache tomcat]. Queries that have several possible interpretations (\emph{intents}), such as [apache], are referred to as \emph{ambiguous} \cite{Song2007}. 

Another possible direction to improve the query suggestion ranking is to diversify a list of proposed suggestions, i.e.\ to avoid unnecessarily redundant candidates. To illustrate,  let us consider a prefix [apache t] and two possible candidate sets: ([apache tomcat], [apache tomcat install], [apache tomcat download]) and ([apache tomcat], [apache tomcat install], [apache territory]). Despite the fact that [apache tomcat download] may be more frequent than [apache territory], its utility in the presence of the two other candidates can be lower than that of [apache territory] and hence the latter set can be more useful to the users.

Often, the diversification and contextualisation approaches are studied independently (e.g., \cite{Santos2010, Sontag2012}). However, if considered independently, they are to some extent contrasting: while diversification implies a broad coverage of intents in order to satisfy as many users as possible, contextualisation aims to cover only the needs of a particular user. In some sense, both approaches constitute possible ways to deal with the lack or presence of the user's intent preference information and a problem arises in how to combine them in a mathematically motivated manner.  

In this paper, we describe a framework that is capable of providing a user whose intentions are clear with highly contextualised queries and a user without context with diversified suggestions. We investigate how both the short-term search context and the query candidates diversification can be combined to provide a user with suggestions that are more useful with fewer typed characters. Our contributions in this paper can be summarised as follows:
\begin{itemize}
    \item A framework to perform contextualised and diversified query suggestions ranking; 
    \item A generative model of user behaviour that is capable of inferring the user's real intent from their short-term context, namely the previously entered query, the documents skipped and clicked, and the discarded query suggestions;
    \item A thorough experimental evaluation of the framework.
\end{itemize}

The rest of this paper is organised as follows. 
After reviewing some related work in Section \ref{sRelated}, we discuss our framework to perform a contextualised and diversified ranking of query suggestions in Section \ref{sCntxDiv}. Next, we describe our proposed approach to model user behaviour (Section \ref{sFrameworkLearning}) and how it can be used to represent the user context (Section \ref{sFrameworkContext}). 
Section~\ref{sXquad} shows how our proposed query suggestion framework is related to an existing state-of-the-art diversification framework. 
The experimental setup of our evaluation is described in Section~\ref{sExperiments}. Finally, we report the results obtained in Section \ref{sResults} and close the paper with some conclusions in Section~\ref{sConclusion}.

\section{Related Work}\label{sRelated}%
Three papers, Bar-Yossef et\ al.~\cite{Bar2011}, Yan et al.~\cite{Yan2011}, and Shokouhi \cite{Shokouhi2013} are the most closely related to our work. Bar-Yossef et al.~proposed a method to contextualise query suggestions, which relies on representing suggestions and context (previous queries) as high-dimensional term-vectors. The ranking of suggestions is then based on the linear combination of the query frequency and the similarity with the previous context. 
Shokouhi \cite{Shokouhi2013} proposed a feature-based machine learning framework to personalise query suggestions. Shokouhi considered the long-term features (e.g.\ age, the user's previous queries) as well as the short-term features (e.g.\ the user's queries in the same session) and demonstrated that his approach outperforms the baseline that ranks suggestions according to their frequency. 
We note three key differences between our work and that of Bar-Yossef et al. \cite{Bar2011} and Shokouhi \cite{Shokouhi2013}: the context we consider also includes those documents that were clicked and skipped; our explicit modelling of user intents makes it possible to diversify the suggested queries while keeping them contextualised. In addition, in contrast to the work of Bar-Yossef et al.\, our method allows us to adjust the trade-off between query popularity and relatedness with the context in a context-dependent manner.  

After submitting a query, some search engines provide the users with query recommendations. 
An approach used by Yan et al.~\cite{Yan2011} to perform query recommendations in the presence of ambiguous que\-ries differs from our work by ignoring the search results the user skipped previously. In addition, despite formulating the user needs in terms of query intents, the diversification of the recommendations was not discussed. Cao et al.~\cite{Cao2008} studied the related problem of contextualising the query recommendations, but without considering the user's document examination behaviour as part of the context or addressing the explicit diversification. The method we use to extract latent user intents from the click and reformulation behaviour can be considered as an analog to the method introduced by Cao et al.~\cite{Cao2009}. However, they only mention the possibility to leverage that information in the query suggestion mechanism. 
Song et al.~\cite{Song2011}  proposed an approach to perform a diversified ranking of query recommendations. The underlying idea behind their work is to promote queries with a high level of novelty with respect to the previously submitted query. Another algorithm to generate diversified query recommendations was introduced by Ma et al.~\cite{Ma2010}. The algorithm leverages a Markov random walk process on a query-URL bipartite graph to infer the most probable recommendation. 
The first difference with our work is that both approaches \cite{Ma2010, Song2011} rely on the query candidate's implicit similarity and dissimilarity without explicitly modelling the possible user intentions. Secondly, the possibility of additional contextualisation of recommendations by the user's click behaviour is not considered.

While our work addresses the diversification of query suggestions, the related task of search result diversification has seen much research in recent years. Various models for search results diversification have been proposed, including IA-Select~\cite{Agrawal2009} and xQuAD~\cite{Santos2010}, which both build a diversified ranking of {\em documents} in an iterative manner, by greedily selecting at each step a document that has a maximal probability to satisfy a user \emph{given that the previously selected documents have failed to satisfy the user}. Since the previously selected documents have failed to satisfy the user, the next selected document should cover intents that are the least covered by those previously selected documents. 
Vallet et al.~\cite{Vallet2012} introduced extensions to diversification frameworks, such as IA-Select \cite{Agrawal2009}, to perform a diversified and personalised ranking of web results. 
Our work differs from that of Vallet et al.\ in the nature of the tackled task (query suggestions vs.\ results diversification) as well as the nature of the considered user behaviour (long-term vs.\ short-term).  


\looseness-1 As can be seen from the related work, little attention has been paid to the contextualisation of query suggestions by means of analysing the user's document examination behaviour, as well as to the problem of combining contextualisation and diversification. 
In the next section we introduce a novel framework that combines the user's short-term query and document examination context and the diversification of the query suggestions in a unified manner, to improve the users' satisfaction with the query suggestions. 

\section{Context and Diversity in Ranking Query Suggestions}\label{sCntxDiv}
Our approach to combine diversification and contextualisation is to reformulate the diversification problem as a special case of contextualisation. We use the underlying idea of IA-Select \cite{Agrawal2009}, where at each iteration, when a new document is selected to be added to the ranked list, the documents previously selected are assumed to have failed to satisfy the user's needs. Similarly, we consider the set of queries already placed in the list of query suggestions to have failed to satisfy the user, and this set of queries forms the \emph{diversification} part of the context. Apart from this, the user's previous behaviour (previously submitted query, documents clicked or skipped) constitutes the \emph{historical} part of the context.

Informally, our framework to build diversified and contextualised suggestions can be recursively defined as follows. 
We consider the session of a user who submitted the query $q_0$, interacted with the search result page and submitted several characters of their next query. 
On the $k+1$th step of building a query suggestions list, we already have selected $k$ suggestion candidates: $Q^k = \{q_1^{1}, ... ,q_1^{k}\}$. The task is to select the next candidate $q_1^{k+1}$ with the highest probability to guess the user's target query, given their historical context $H$, and $Q^k$ as a diversification context. We denote the full context available at step $k$ of the algorithm as $T_k := H \cup Q^k$. Then, the next query suggestion $q^{k+1}_1$ is greedily selected with the highest probability to be submitted by the user given the current context $T_k$, $P(q^k_1|T_k)$. After finding the suggestion candidate with the highest probability, it is included into $Q^k$ as $q_1^{k+1}$: $Q^{k+1} := Q^{k} \cup \{q^{k+1}_1\}$. We repeat the procedure until the required number of suggestions is selected. 

As the above algorithm can encapsulate different forms of information, it constitutes a framework that can generate different query suggestion lists.  In the remainder of this section, we describe a context-aware method to estimate the probability $P(\bar q_1|T_k)$\footnote{
 For clarity, in the following we use the simpler notation $\bar q_1$ instead of $q^{k+1}_1$ for a candidate query suggestion identified at step $k+1$. The query the user actually submits is referred to as $q_1$.}
for ranking candidate query suggestions given the current context $T_k$.


 


Since we are considering the problem of ranking query suggestions after observing a part of the user's session, it is natural to assume that the set of possible user intentions (search tasks) $I$ coincides with a set of possible interpretations of the previous query $q_0$, i.e. $I = I(q_0)$. This allows us to deal with the small set of intentions associated with $q_0$, instead of the potentially huge space of all possible search intents users may have. On the other hand, we lose the ability to diversify the candidate queries that are not represented by the intents of $q_0$, which may reduce the usefulness of diversification.
A previous search task can be completely unrelated to a new query $q_1$ if a user has satisfied his/her information need and is starting a new search task with the new query.
Following Ozertem et al.~\cite{Ozertem2012}, we use the notion of \emph{search task continuation} to account for this effect. 
Let us introduce an indicator variable $c$, which is equal to $1$ if the user's previous task is continued, and denote the probability of the continuation given the user's context as $P(c = 1|T_k)$.


We assume that before starting to submit the next query, the user can be in one of the $|I| + 1$ states: either the user is satisfied with the results and is not continuing their previous search task (one state); or the user is dissatisfied with the results and continues the task ($|I|$ intent states). At each step, we update the full context $T_k$ and consequently our beliefs of the user's state. After that, we find a suggestion candidate with the highest expected probability to be submitted by the user. To formalise this idea, we expand the expected probability of submitting $\bar q_1$ given the current context $T_k$:
\begin{equation}
P(\bar q_1|T_k) = P(\bar q_1, c = 1|T_k) + P(\bar q_1, c = 0|T_k)
\end{equation}
The first term corresponds to the probability of the user submitting $\bar q_1$ and continuing the previous search task. The second term equals to the probability of submitting $\bar q_1$ while starting a new search task. We assume that a user who is not going to continue their current search task issues a query $\bar q_1$ with probability $P_g(\bar q_1)$,\footnote{Except for $P_g(\bar q_1)$, all probabilities in Sections \ref{sCntxDiv} and \ref{sFrameworkLearning} are conditioned by $q_0$ - we omit $q_0$ to simplify the notation.} which is close to the observed frequency of the query $\bar q_1$ in a query log.
Taking that into account, we can drop the query's dependency on the context in the second term, and as a consequence, we can re-write the previous expression as:
\begin{equation}
\label{eq:General}
P(\bar q_1|T_k) = P_g(\bar q_1) P(c = 0 | T_k) + P(\bar q_1|T_k, c = 1) P(c = 1|T_k)
\end{equation}

\looseness -1 Since the estimates for $P_g(\bar q_1)$ can be found directly by counting query occurrences in the query log, and are independent of the user's context, we focus on estimating our belief that the user continues their task (i.e. $P(c = 1|T_k)$) and the probability that the user submits $\bar q_1$ given he/she continues the search task, $P(\bar q_1|T_k, c = 1)$. To estimate the latter probability, we assume that context $T_k$ influences these probabilities only by affecting the distribution of user intents.
To leverage this assumption, we firstly factorise $P(\bar q_1|T_k, c = 1)$ over the possible intents $I(q_0)$:
\begin{equation}
\label{eq:pqContext}
P(\bar q_1|T_k, c = 1) = \sum_{i \in I(q_0)} P(\bar q_1 | T_k, c = 1, i) P(i | T_k, c = 1)
\end{equation}
Under the above assumption, the probability of query $\bar q_1$ is independent from the user's context given the user's intent, thus we can drop the conditioning on the context from the first term:
\begin{equation}
\label{eq:B}
P(\bar q_1|T_k, c = 1) = \sum_{i \in I(q_0)} P(\bar q_1 | c = 1, i) P(i | T_k, c = 1)
\end{equation}

\noindent Using Bayes' rule, $P(i | T_k, c = 1)$ can be estimated as follows:
\begin{equation}
\label{eq:D}
P(i | T_k, c = 1) = \frac{P(T_k | c = 1, i) P(c = 1, i)}{\sum_{j \in I} P(T_k | c = 1, j) P(c = 1, j)}
\end{equation}

As a next step, we obtain the following expression to estimate the task continuation probability $P(c = 1 | T_k)$:\
\begin{equation}
\label{eq:A}
P(c = 1 | T_k) = \frac{\sum_{j \in I} P(T_k | c = 1, j) P(c = 1, j)}{\sum_{j \in I, t \in \{0,1\}} P(T_k | c = t, j) P(c = t, j)}
\end{equation}
The probability $P(c = 0|T_k)$ can also be similarly calculated.

Finally, $P(\bar q_1|T_k)$ can be estimated by combining Equations (\ref{eq:B}), (\ref{eq:D}) and (\ref{eq:A}) and putting these into Equation (\ref{eq:General}). The obtained expression can be used for query ranking with various representations of the context $T_k$. Only the probabilities of observing the context $T_k$ ($P(T_k| c = 0, i)$ and $P(T_k | c = 1, i)$) depend on the context representation. 

We have thus far defined our framework. In the next two sections, we discuss two possible instantiations of the framework and  discuss how $P(c, i)$, $P(\bar q_1 | c = 1, i)$, and the context representation parameters can be learned from a query log.

\section{Intent Modelling}\label{sFrameworkLearning}
In this section, we introduce a generative approach to model the user behaviour. This model provides us with the means to represent different forms of the user context, as we will discuss in Section \ref{sFrameworkContext}. This representation is further used in the above proposed query suggestion framework.

A part of the session, which starts with the user submitting a query and finishes with a query reformulation $q_1$, is further referred to as \emph{an interaction}. Let us consider a population of user interactions all starting with the query $q_0$, $O(q_0)$. We assume that $O(q_0)$ is generated from a mixture of models, with each mixture component corresponding to an intent $i$ from the family $I(q_0)$. With each interaction $o \in O(q_0)$, we associate two latent variables: the intent $i(o)$ the user had while submitting $q_0$ and a binary variable $c(o)$ which is equal to 1, if the next query $q_1$ belongs to the same search task as $q_0$ and is equal to 0, if the user decided to switch to another task.

\begin{table}[tb]
    \caption{Notations used}
\resizebox{80mm}{!}{
    \label{table:notation}
    \begin{tabular}{ ll }
    \toprule
    Symbol & Description \\
    \midrule
	$E_j$ & was the $j$th document examined? \\
    $K_j$ & was the $j$th document clicked? \\
	$A_j$ & was the user attracted by the $j$th document? \\
	$S_j$ & was the user satisfied with the $j$th document? \\
    $N_j$ & was the next query submitted after examination\\
          & of the $j$th document? \\
	$R$   & a query submitted by the user at the end of \\
        & interaction \\
	$C$   & a latent binary variable denoting if the user  \\
          & continues his/her search task while submitting $R$ \\
    $I$  & an intent latent variable \\
        \bottomrule
    
    \end{tabular}
}
\end{table}

Each mixture component describes a model inspired by the Simplified DBN \cite{DBN} click model and a unigram language model over possible query reformulations.
Our model of user behaviour assumes that after submitting $q_0$, a user with intent $i \in I(q_0)$ examines results from top to bottom, one at a time. An examined document $d$ attracts a user's click with probability $a_{d,i}$ and satisfies
the user after clicking with probability $s_{d,i}$. If the user is satisfied with the last result clicked then the next submitted query $q_1$ is unrelated to the previous search task and its terms $t \in q_1$ are distributed according to the background unigram language model of the whole query stream $P^{lm}_{g}(t)$. If the user is not satisfied, then the user continues to examine documents until they find a satisfying document or after examining all the query results they submit a new query with an intent-dependent term distribution $P^{lm}(t|c = 1, i)$.

In other words, the model assumes that a term $t$ of $q_1$ is generated from a mixture of $|I| + 1$ components and a user's satisfaction with the last clicked document determines which component will be used to generate it: a non-satisfied user submits a query with terms generated from $P^{lm}(t|c = 1, i)$ while a satisfied user generates the terms of the next query from the distribution  $P^{lm}_g(t)$. A user who cannot find a satisfying document examines all query results. Due to these intent-dependent click and language models, interactions with similar click/skip patterns or reformulations tend to be associated with the same component of the mixture.

\begin{figure}[tb]
\centering
\caption{A graphical model of the user behaviour. Grey circles correspond to observed variables.}
\resizebox{65mm}{!}{
\input{graphical_model.tikz}
}
\label{ref:model}
\end{figure}
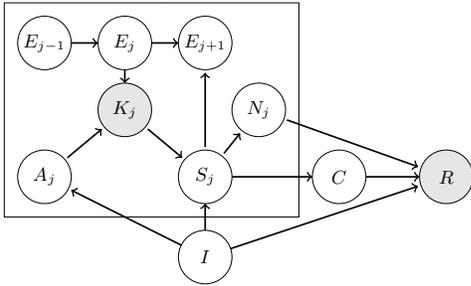

The underlying graphical model is depicted in Figure \ref{ref:model} and, for a given document position $j$, it uses the random variables described in Table \ref{table:notation}.
Denoting the document on the $j$th position as $d_j$, the model can be described by means of the following equations:
\begin{subequations}
\begin{align}
\label{model:a} E_1 = 1\\
\label{model:b} E_j = 0 \Rightarrow E_{j+1} = 0\\
\label{model:c} A_j = 1, E_j = 1 \Leftrightarrow K_j = 1\\
\label{model:d} P(A_j| I = i) = a_{d_j,i}\\
\label{model:e} P(S_j|K_j = 1, I = i) =  s_{d_j, i}\\
\label{model:f} K_j = 0 \Rightarrow S_j = 0 \\
\label{model:g} \exists j ~ S_j = 1 \Rightarrow E_{j+1} = 0, C = 0, N_j = 1\\
\label{model:h} S_j = 0, E_j = 1, j < J \Rightarrow E_{j + 1} = 1, N_j = 0\\
\label{model:i} \forall j \le J ~ S_j = 0 \Rightarrow C = 1, N_j = 1\\
\label{model:j} P(R | C = 0, N_j = 1) = \prod_{t \in Q} P^{lm}_g(t)\\
\label{model:k} P(R | C = 1, I = i, N_j = 1) = \prod_{t \in Q} P^{lm}(t|c = 1, i)
\end{align}
\end{subequations}
Indeed, the above equations describe the following constraints on the model: The first document is always examined \eqref{model:a}; Documents are examined  sequentially \eqref{model:b}; When an examined document is attractive, the user will click it \eqref{model:c}; the probability of attracting the user and the probability of satisfying the user are the document parameters, conditioned on intent \eqref{model:d} \& \eqref{model:e}; An unclicked document cannot satisfy a user \eqref{model:f}; The examination of the ranked document list terminates when the user is satisfied, meaning that the search task is not continued and the user submits a new query \eqref{model:g};  If the user is not satisfied, then the examination proceeds down the ranked list, as far as rank $J$ \eqref{model:h}; If the user is not satisfied with the top $J$ documents, then the user continues their search task with a new query \eqref{model:i}; A new query for a new search task of a satisfied user is drawn with the likelihood as given by the query log \eqref{model:j}; However, for the next query in a continuing search task, this probability is conditioned on the intent of the user \eqref{model:k}.

The maximum a posteriori (MAP) estimates of the model parameters $\theta = (a_{d,i}, s_{d,i}, P(w | i) )$ as well as the $P(c, i)$ distribution are found from the available query log data by means of an Expectation-Maximisation (EM) procedure:
$$
\hat \theta(q_0) = argmax_{\theta} \sum_{o \in O(q_0)} \log P_\theta(o) + \log P_{prior}(\theta)
$$
where $P_\theta(o)$ denotes the probability of an observed interaction $o$ given the model parameters $\theta$.
Following \cite{DBN} and \cite{SLM}, we impose Beta and Dirichlet priors on the click ($a_{d,i}$, $s_{d,i}$) and language model  ($P^{lm}$, $P_g^{lm}$) parameters, respectively.
The probability $P(q_1|c = 1, i)$ is obtained by performing a single Expectation-like step over the interactions ending with $q_1$:
$$
P(q_1|c = 1, i) = \frac{ \sum_{o \in O(q_0, q_1)} P(c = 1, i|o) } { \sum_{o \in I(q_0)} P(c = 1, i|o) }
$$
where $P(c = 1, i| o)$ are the responsibility values obtained on the Expectation step and $O(q_0, q_1)$ is a set of interactions starting with $q_0$ and ending with $q_1$.

\section{Context Representation}\label{sFrameworkContext}
Recall that within our framework, we consider the context $T_k$ for a suggestion $\bar q_1$ at the $k+1$th step to be defined as the unselected queries $Q^k$ and the historical context. In the following, we define two possible context representations for the historical part $H$ of the context $T_k$. The first representation includes the previous query $q_0$ only, while the second includes not only the query, but also documents that the user has clicked or skipped during the session.

\textbf{Query-only history} 
Once the query-only historical context $H$ is considered, 
our current belief in the user's state is determined by the query-dependent intent probabilities $P(c, i)$ and the diversification part of the context $Q^k$. Since we assume that the probability of the suggestion $r \in Q^k$ to be submitted by the user is independent from the context, given the user's intent $i$, the probability that all $k$ previously ranked suggestions have failed to guess the user's target query given the user's intent $i$ is equal to:
\begin{equation}
\label{eq:E}
P(T_k | c = 1, i) = P(Q^k| c = 1, i) = \prod_{r \in Q^k} \left[ 1 - P(r|i, c = 1) \right]
\end{equation}

On the other hand, if the user decided not to continue the task, then the assumption about the previously selected candidates leads to the following representation of the probability of the context $T_k$:

\begin{equation}
\label{eq:F}
P(T_k | c = 0, i) = P(Q^k| c = 0) = \prod_{r \in Q^k} \left[ 1 - P_g(r) \right]
\end{equation}

We expect the query-only context to be useful since the knowledge of the previous query should dramatically reduce the space of the user's possible intentions.

\textbf{Query, clicks and skips as history}
A more detailed search context includes not only the previous query, but also the clicked and skipped documents for that query.  
Taking the user intent model proposed in Section \ref{sFrameworkLearning} into account and assuming that the diversification and historical contexts are independent given the user's intent $i$ and the search task continuation indicator $c$, the probability of context $P(T_k | c = 0, i)$ can be expressed as follows:
\begin{equation}\label{eq:G0}
P(T_k | c = 0, i) = P(H|c = 0, i)P(Q^k|c = 0)  
\end{equation}
Similarly, 
\begin{equation}\label{eq:G1}
P(T_k | c = 1, i) = P(H|c = 1, i)P(Q^k|c = 1, i)  
\end{equation}
with $P(Q^k| c = 0)$ and $P(Q^k|c = 1, i)$ being defined by Equations (\ref{eq:E}) \& (\ref{eq:F}), respectively.

As in Section \ref{sFrameworkLearning}, $d_j$ denotes the document in the $j$th position and $K_j$ is a binary variable representing that $d_j$ was clicked.
According to the user model described in Section \ref{sFrameworkLearning}, the probability of observing the historical part of the context is equal to:
\begin{equation}
\label{eq:G2}
\begin{array}{lll}
P(H|i, c = 0) = a_{l,i} s_{l,i} & \prod_{j=1}^{l-1} a_{d_j,i}^{K_j}( 1 - s_{d_j,i})^{K_j} \cdot \\
 &     \cdot  (1 - a_{d_j,i})^{1-K_j}
\end{array}
\end{equation}
\begin{equation}
\label{eq:G3}
\begin{array}{lll}
P(H|i, c = 1) = & \prod_{j=1}^{J} a_{d,i}^{K_j}( 1 - s_{d,i})^{K_j} \cdot \\
 &     \cdot      (1 - a_{d,i})^{1-K_j}
\end{array}
\end{equation}
where $J$ is the number of results on a search page. 
Putting Equations (\ref{eq:G0}), (\ref{eq:G1}), (\ref{eq:G2}) \& (\ref{eq:G3}) into Equation (\ref{eq:A}), we can find $P(\bar q_1|T_k)$ with the search context represented by the document's clicks and skips. The values of $P(c, i)$, $a_{d,i}$, $s_{d,i}$ and $P(r|c = 1, i)$ are estimated from a session log using the EM procedure, as we discussed in Section \ref{sFrameworkLearning}.

\section{Relation to xQuAD}\label{sXquad}
xQuAD \cite{Santos2010} is one of the state-of-the-art frameworks for web search results diversification -- as illustrated by its top performances in recent TREC Web track diversity evaluations (e.g.~\cite{clarke2012trec}). The framework builds the diversified result list of documents in a greedy manner, at each step selecting a document that maximizes the following objective function:
\begin{equation}
\label{eq:xquad}
f(d) = (1 - \lambda) P(d|q) + \lambda \sum_i P(i|q) P(d|i) \prod_{\bar d \in S}\left(1 - P(\bar d|i) \right)
\end{equation}
where $P(d|q)$ is the document's relevance to the query $q$, $P(i|q)$ is the query intent distribution, $P(d|i)$ denotes the document's relevance with respect to an intent $i$ and $\lambda$ is a free parameter.

Bearing in mind that in a query suggestion ranking, the query candidates take the place of documents, and comparing Equation (\ref{eq:xquad}) with Equations (\ref{eq:General}), (\ref{eq:pqContext}) \& (\ref{eq:E}), we can find several noticeable similarities:  $\lambda$ corresponds to the probability of search task continuation $P(c = 1 | T_k)$, the document's non-diversified relevance $P(d|q)$ corresponds to $P_g(\bar q_1)$, and the query's intent distribution $P(i|q)$ corresponds to $P(i|T_k, c = 1)$. Thus the proposed framework can be seen as a contextualised extension of xQuAD to the query suggestions domain. However, apart from the ability to leverage short-term context dependencies to infer the intent distribution, the two frameworks differ in the way the relevance-diversity (or popularity-relatedness) trade-off is addressed: while xQuAD and its personalised modification \cite{Vallet2012} rely on a model parameter $\lambda$ to adjust the trade-off, our proposed framework uses a context to set the probability of the search task continuation $P(c = 1|T_k)$. 

From that point of view, a work by Santos et al. \cite{Santos2011b} discusses similar ideas to some extent. Indeed,  Santos et al. propose a feature-based machine learning approach to predict if a query aspect implies a navigational or informational search task and leverages that information in the diversification algorithm. 

The preceding sections have defined our proposed framework, introduced a Bayesian model for query and document contexts, and have shown how it relates to an existing document diversification framework. In the following, we define the experimental setup that we use to empirically evaluate our framework's behaviour within a large commercial search engine.

\section{Experimental Setup}\label{sExperiments}
Our empirical study has the following goals. Firstly, we investigate if the proposed framework to perform a diversified and contextualised ranking leads to improvements over a basic baseline algorithm that ranks query candidates according to their frequency. Secondly, we compare the performance of different possible combinations of historical and diversification contexts, so as to determine which part of the context, historical or diversification, leads to better improvements. 

To address these goals, we experiment with the baseline ranking and four different variations of the proposed ranking framework: 
(a) Ranking contextualised by only the previous query entered by the user; (b) Ranking contextualised by the previous query and the documents clicked and skipped; (c) Ranking diversified and contextualised by the previous query; (d) Ranking diversified and contextualised by the previous query and the clicked and skipped documents.

The case of contextualisation by the previous query only (a) (with $T_k = q_0$) corresponds to ranking query candidates according to the probability of their generation from a mixture of the background query stream distribution $P_g(\bar q_1)$ and the distribution of query reformulations with the user's task unchanged $P(\bar q_1 | c = 1)$:
$$
P(\bar q_1|q_0) = P(c = 0 | q_0) P_g(\bar q_1) + P(c = 1|q_0) P(\bar q_1 | c = 1, q_0)
$$

The mixing coefficients $P(c = 0 | q_0)$ and $P(c = 1 | q_0)$ are query-dependent and can be found by marginalising $i$ from $P(c, i | q_0)$ known from the model learning step.



\textbf{Dataset}
 The dataset that we use in our evaluation experiments consists of training and test parts. The training part was generated from Yandex' query log over a period from June, 1 to August, 28, 2012. The following two weeks were used to create the test set. We split the user actions into interactions (a part of session between two queries - as defined in Section~\ref{sFrameworkLearning}) by five minutes of inactivity. A production query suggestions mechanism with near-duplicate queries removed was used to calculate $P_g(q)$ using the training dataset.
In order to avoid sparsity while learning the model parameters, we filter out all queries with less than 400 interactions observed during the training period.  The test set contains only interactions starting with a query in the training set. Some descriptive statistics of the datasets can be found in Table \ref{tab:stat}.

\begin{table}
\centering
\caption{Dataset statistics}
\label{tab:stat}
\begin{tabular}{ ccc }
\toprule
& Train & Test \\
\midrule
time period & 3 months  & 2 weeks\\
\#interactions $(o)$& 19M  & 960k\\
\#unique reformulations $(q_1)$ & 11M & 340k\\
\#unique queries ($q_0$) & 17k & 7k\\
\bottomrule
\end{tabular}
\end{table}

\textbf{Estimating The Model Parameters}
All model parameters - namely $P(c, i)$, $a_{d,i}$, $s_{d,i}$ and $P(q|c = 1, i)$ are estimated from the training set using the Deterministic Annealing modification of the EM algorithm \cite{Ueda1998}, as initial experiments found this to obtain the highest performance.  A hold-out subset of the training data is used to adjust the parameters of the Beta and Dirichlet priors. While learning the language models of reformulations, the queries are lemmatised and stopwords are removed.  
The EM procedure used to estimate the model parameters is only able to find a local optima of the objective, thus the results obtained are sensitive to the initialisation of the latent intent variables. Moreover, in order to run the EM optimisation, the number of mixture components (user intents, $I$) should be determined beforehand. A variety of approaches to address these requirements of the EM procedure have been proposed, e.g. Figueiredo et al.~\cite{UnsupEM2002} leverage the minimum description length principle to automatically adjust the number of mixture components. However, the optimal choice of initialisation parameters is out of the scope of the paper and, for this reason, we use an entity-based web search intent-mining algorithm both to set the number of intents as well as to initialise the latent intent variables:
an interaction is assigned to an intent which is the most likely connected with the last clicked document in the interaction. The number of intents for each query is set equal to the number of intents identified by the web search diversification algorithm. 
The intent-mining algorithm works in two steps. At the first step, the system analyses the users' queries and identifies entities occurring (films, books, etc.) in the queries. This process is weakly supervised and relies on query template mining. At the next step, each entity is classified into one or more manually predefined categories, based on \emph{category indicators} that frequently co-occur with the entity in the users' queries.  These steps are related to the algorithm described by Pa\c{s}ca \cite{Pasca2007}.
The category indicators and a set of possible intentions for a category are extracted from the query log in a semi-automated manner.
For instance, the query [casablanca] will be classified both into the ``city" and ``film" categories, with ``film" category having such intents as ``buy a dvd", or ``reviews". 
Since the underlying web result diversification algorithm is used only for the EM initialisation, the proposed learning algorithm does not depend on its implementation and can be used with any diversification algorithm.  In fact, given some reasonable initialisation, the algorithm itself is capable to extract latent intents from a query log. Due to the EM initialisation scheme, all queries without intents known from the underlying web search diversification algorithm are removed from the dataset. 
In order to speed up the learning process, we restrict each query to have no more than 5,000 associated interactions, uniformly sampling the required number of interactions for highly frequent queries. The optimisation is terminated either after performing 75 iterations or when the difference in log-likelihood between two consecutive iterations is less than $0.005$. Since the intent model mixtures are trained on a per-query basis, the learning process can be easily parallelised within a MapReduce framework \cite{Dean2008}.

\textbf{Evaluation}
Unfortunately, there is no metric for evaluating query suggestion quality that is commonly accepted in the literature.
Shokouhi et al.~\cite{Shokouhi2012} evaluated the quality of suggestions for a given prefix by the reciprocal rank of the most popular result, and the Spearman correlation between the predicted and ground-truth ranks of the queries. These metrics were averaged over a set of test prefixes. They consider a query as relevant if it is top-ranked according to the ground-truth query frequencies, Strizhevskaya et al.~\cite{Alisa2012} reported P@3, AP@3 and nDCG@3 averaged over the observed prefixes. 
We use the same session log-based scenario as Bar-Yossef et al.~\cite{Bar2011}, with two minor changes. 
Since building the diversified ranking list is of quadratic computational complexity with respect to the number of suggestion candidates, we perform diversification in two steps. At the first step, we use the corresponding non-diversified contextualised ranking to find 100 top scored candidates. Next, we perform the diversified re-ranking of these candidates. As a result of this scheme, weighting the scores by the number of candidates becomes less justified. Further, as the length of the query suggestion lists is usually no longer than 10, cut-off levels higher than 10 do not reflect the user's actual experience. For all these reasons, we report mean reciprocal rank (MRR) at cut-off level 10 in our experiments. Following \cite{Shokouhi2012}, we use a prefix of length 3 to filter the query candidates. Recently proposed metrics such as \emph{pSaved/eSaved} \cite{Kharitonov2013Sigir} also assume the query log-based evaluation approach and can be used as well. 

In a web search setting, the evaluation of diversification algorithms usually implies labelling documents manually by judges and calculating intent-aware quality metrics, such as ERR-IA \cite{Chapelle2011}. In contrast, we measure diversification success as the ability to rank a suggestion candidate preferred by a user higher, using a query log as an evidence of that preference and MRR as a metric. We believe that a query log-based approach better reflects the user experience. However, it has some drawbacks, e.g.\ it is assumed that only the query the user submitted can satisfy her, even if there are semantically similar queries ranked higher which the user might have not seen in her session. Thus we consider the benefit obtained from contextualisation and diversification of the query suggestions in our experiments as a lower bound of the real improvement.

Overall, to the best of our knowledge, our work is the first to evaluate the effects of diversification using a query log-based offline approach.
     \begin{table*}[tb]
    \centering
    \caption{Relative improvements in $MRR@10$ over the baseline after the user submits the first three characters of the query. $|q_1|$ is the length of the second query in words. 
}
        \begin{tabular}{ cccccr}
        \toprule
                  &  \#interactions    & QCntx     & QCntxDiv &          FCntx & FCntxDiv \\
    \midrule
         $|q_1| > 0$ & 960k     & +0.302 & +0.304 & +0.307  &  +0.307 \\
         $|q_1| > 1$ & 715k     & +0.509  & +0.513 & +0.518 & +0.519 \\
         $|q_1| > 2$  & 412k     &  +0.811 & +0.818 & +0.825 & +0.826 \\
         $|q_1| > 3$  & 224k     & +0.977 & +0.987 & +1.001 & +1.003 \\
         \bottomrule
        \end{tabular}
    \label{tab:resultsLength}
    \end{table*}

\section{Results and Discussion}\label{sResults}
\looseness-1 In this section, we report the evaluated quality of different combinations of contexts. Moreover, as longer queries are harder to predict given the first three characters, we obtain additional insights into the framework's performance by varying the length of $|q_1|$ in the experiments. The results are presented in Table \ref{tab:resultsLength}. We use the following abbreviations: QCntx corresponds to the ranking with the previous query as a context; FCntx corresponds to the non-diversified ranking with the previous query, clicks and skips as a context; QCntxDiv and FCntxDiv correspond to the versions of QCntx and FCntx with the diversification context added.

Due to the proprietary nature of the system, we report only relative improvements  over the baseline (e.g.\ $+0.302$ denotes a 30.2\% relative improvement). All pairwise differences of these improvements (i.e., for any two cells on a single row with a non-zero difference reported) are statistically significant according to the paired t-test ($p < 0.005$).

\begin{table*}
    \centering
    \caption{Relative improvements in $MRR@10$ over the baseline after the user submits the first three characters of the query. $isPrefix$ is a binary variable representing if the second query $q_1$ contains the first query $q_0$ as a prefix. $|q_1|$ is the length of the second query in words.
}
        \begin{tabular}{cccccr}
        \toprule
                       & \#interactions & QCntx     & QCntxDiv &          FCntx & FCntxDiv  \\
        \midrule
         $isPrefix(q_0,q_1) = 0, |q_1| > 0$ & 730k  & +0.201 & +0.203 & +0.201 & +0.202 \\
         $isPrefix(q_0,q_1) = 0, |q_1| > 1$ & 515k  & +0.317   & +0.322 & +0.316 & +0.317 \\ 
         $isPrefix(q_0, q_1) = 1, |q_1| > 0$ & 230k & +0.770 & +0.772 & +0.798  &  +0.798 \\
         $isPrefix(q_0, q_1) = 1, |q_1| > 1$ & 200k  & +1.268 & +1.269 & +1.315  & +1.315 \\ 
         \bottomrule
        \end{tabular}
    \label{tab:resultsPrefix}
    \end{table*}

 On analysing Table \ref{tab:resultsLength}, we observe that contextualisation leads to a considerable improvement over a basic query suggestion approach that simply ranks candidates by their frequency. This agrees with the results reported in \cite{Bar2011}, however the values of the improvements are not directly comparable.

A noteworthy observation is that as the length of the second query in words grows, the improvement from the context increases.  This seems reasonable, since the longer queries are harder to predict by the baseline approach given the three-character prefix. On the other hand, longer queries are related to the reformulation behaviour when a user is dissatisfied with the first query and tries to specify it using additional keywords. In this scenario, the user's search task continues and hence contextualisation is beneficial. 

For all considered subset of queries, adding the document examination context leads to further increases in the contextualisation performance.  Moreover, the relative improvement from the additional contextual information grows as the query length grows. We believe that a richer context allows the framework to derive the possibility of the task continuation and the user's intent with higher confidences, thus resulting in better results on sessions with reformulation behaviour.

In order to support the idea that the query specification affects the benefit of contextualisation, we additionally consider the case of $q_0$ being a prefix of $q_1$ and report the results in Table \ref{tab:resultsPrefix}. Indeed, we can see that contextualisation exhibits considerable performance improvements for interactions when the second query specifies the first one. In addition, for such interactions, the relative gain from adding click behaviour reaches its maximum. Our intuition behind this observation is that given the richer search context, the framework is able to contextualise the candidates more aggressively and since for that interaction the second query is indeed related to the previous task, this results in significant improvements. On the other hand, the contextualisation is useful for interactions where the second query does not contain the first one as a prefix, demonstrating that the relatedness of queries goes beyond simple prefix-similarity.

In our experiments, the benefit of the diversification context is less marked than the benefit of the historical context (though statistically significant), especially when we have more evidence about the user intent (FCntx). This observation makes sense, since diversification in its nature is loosely a method to address our uncertainty in the user's search task, while contextualisation leverages information to predict the task and rank suggestion candidates accordingly.
An interesting observation from Table \ref{tab:resultsPrefix} is that if the user does not continue her task (which is not known at the time of ranking suggestion candidates) then the improvement from adding the diversification context is higher than in the opposite case. This observation supports the idea of diversification as a tool to mitigate uncertainty in the user's intentions, thus being more useful for users not continuing their search tasks and less useful for user continuing their search tasks.

On the other hand, this is not the case when ranking with a richer context, FCntx, possibly due the fact that the click behaviour context allows the framework to infer the continuation and the user's intent with a higher level of confidence. 


\looseness-1 Overall, our results support the benefit of enriching the search context with the document examination behaviour as an approach to improve the user's satisfaction with the query suggestion mechanism. Further adding the diversification context does not hurt performance and results in small, though statistically significant improvements in some cases.

To conclude, we find that our proposed framework is able to perform an effective contextualised ranking of query suggestions, 
by handling ambiguity in the user's task.

\section{Conclusions and Future work}\label{sConclusion}
In this paper, we presented a novel framework that performs a contextualised ranking of query suggestions, where the context encompasses the user's previous query, the documents previously clicked and skipped, and the query suggestions already examined. In contrast to the approaches previously discussed in the literature, the proposed framework is capable to combine contextualisation and diversification in a uniform manner. In order to do  so, the diversity requirement is represented as an intrinsic part of the user's search context.

We experimented with two types of historical evidence for the search context: the first one contains the previous query only, while the second additionally contains the documents clicked and skipped during the user's interaction for the previous query. In order to infer the user's intentions from their examination behaviour, we described an approach to model the user behaviour as a mixture of intent models.
Our empirical study using a 3.5 month query log encapsulating about 20M interactions demonstrates that the proposed framework ranks query suggestions better than a baseline approach (approximately a $30\%$ relative improvement on the test set).
Our results also show that enriching the search context with a finer-grained representation of user behaviour leads to further improvements in the suggestion ranking quality.  Indeed, the FCntxDiv ranking with the richest context considered (the user's previous query, document examination history, and diversification context) exhibits the top performance on all the considered subsets of queries and attains a $100.3\%$ relative improvement over the baseline in one of the experiments.
As a possible direction of future work, the same approach can be used to contextualise and diversify web search results. 

\bibliographystyle{abbrv}
\bibliography{bib}  
\end{document}

%% file: graphical_model.tikz
\begin{tikzpicture}[scale=0.9]
   	\draw[fill=white] (1,3) circle (0.5) node{$E_{j-1}$};
	\draw[->,thick] (1.5,3)--(2,3);
	
   	\draw[fill=white] (2.5,3) circle (0.5) node{$E_j$};
	\draw[->,thick] (3,3)--(3.5,3);
	\draw[->,thick] (2.5,2.5)--(2.5,2.25);

   	\draw[fill=white] (4,3) circle (0.5) node{$E_{j+1}$};
	
	\draw[fill=white] (5,1.75) circle (0.5) node{$N_j$};
	\draw[shorten <=0.5cm, shorten >=0.5cm,->,thick] (5,1.75)--(8.5,0.5);
	
   	\draw[fill=gray!20] (2.5,1.75) circle (0.5) node{$K_j$};

	\draw[shorten <=0.5cm, shorten >=0.5cm,->,thick] (1,0.5)--(2.5,1.75);
   	\draw[fill=white] (1,0.5) circle (0.5) node{$A_j$};

	\draw[shorten <=0.5cm, shorten >=0.5cm,<-,thick] (4,0.5)--(2.5,1.75);	
	\draw[->,thick] (4.5,0.5)--(6,0.5); 
	\draw[shorten <= 0.5cm, shorten >=0.5cm, ->,thick] (4.0,0.5)--(5.0,1.75); 
	\draw[shorten <= 0.5cm, shorten >=0.5cm, ->,thick] (4.0,0.5)--(4,3); 
   	\draw[fill=white] (4,0.5) circle (0.5) node{$S_j$};
	
    \draw (0.25,-0.25) rectangle (5.75,3.75);	
	
   	\draw[fill=white] (6.5,0.5) circle (0.5) node{$C$};			
	\draw[->,thick] (7,0.5)--(8,0.5);

   	\draw[fill=gray!20] (8.5,0.5) circle (0.5) node{$R$};
	
	\draw[->,thick] (4,-0.5)--(4,0);
	\draw[shorten >=0.5cm,->,thick] (4,-1)--(1,0.5);
	\draw[shorten >=0.5cm,->,thick] (4,-1)--(8.5,0.5);		
   	\draw[fill=white] (4,-1) circle (0.5) node{$I$};			
\end{tikzpicture}